\documentclass[10pt,conference,compsocconf]{IEEEtran}
\usepackage{amsmath,amssymb}
\usepackage{url}
\usepackage{graphicx}
\begin{document}
\title{Solving $k$-Nearest Neighbor Problem on Multiple Graphics Processors}
\author{\IEEEauthorblockN{Kimikazu Kato and Tikara Hosino}
\IEEEauthorblockA{Nihon Unisys, Ltd.}}
\sloppy
\maketitle
\begin{abstract}
A recommendation system is a software system to predict customers' unknown
 preferences from known preferences.
In a recommendation system, customers' preferences are
encoded into vectors, and finding the nearest vectors to each vector
is an essential part. This vector-searching part of the problem is called a
$k$-nearest neighbor problem. We give an effective algorithm to solve
 this problem on multiple graphics processor units (GPUs).

Our algorithm consists of two parts: an $N$-body
 problem and a partial sort. For a algorithm of the $N$-body problem, we
 applied the idea of a known algorithm for the $N$-body problem in physics,
 although another trick is need to overcome the problem of small sized
 shared memory. For the partial sort, we give a novel GPU algorithm
 which is effective for small $k$. In our partial sort algorithm, a heap
 is accessed in parallel by threads with a low cost of synchronization.
 Both of these two parts of our algorithm utilize maximal power of
 coalesced memory access, so that a full bandwidth is achieved.

 By an experiment, we show that when the size of the problem is large, an
 implementation of the algorithm on two GPUs runs more than 330 times
 faster than a single core implementation on a latest CPU. We also show
 that our algorithm scales well with respect to the number of GPUs.
\end{abstract}

\newcommand{\gsize}{\mathtt{GSIZE}}
\newcommand{\bs}{\mathtt{BSIZE}}
\newcommand{\ndev}{\mathtt{nDevices}}
\newcommand{\cx}{\mathtt{C1}}
\newcommand{\cy}{\mathtt{C2}}

\section{Introduction}
A recommendation system is a software system which utilizes known
customers' preferences to predict unknown preferences. It is widely used
in Internet-based retail shops and other service providers, such as
Amazon.com~\cite{amazon} for example. In a recommendation system,
customers' preferences or buying patterns for items are encoded into
vectors and finding nearest vectors is an essential part of its
computation. This vector-finding part is called a $k$-nearest neighbor
problem. We give an effective GPU algorithm to solve this problem.

Generally a recommendation system deals with large samples and large
dimensions, as person$\times$item for example.  In such a case, the
dimensionality reduction method such as singular value decomposition or
latent Dirichlet allocation has been widely used
\cite{brand03,blei03,das07}.  As the result of the reduction, the
problem becomes the $k$-nearest neighbor search for a moderate dimension.
However, the effect of the sample size $n$ is $O(n^2)$ and it is a
computational burden. Therefore some approximation has been considered
to be necessary~\cite{indyk98}. This paper indicates that strict
computation in practical time is possible. Our target size for $n$ is
$\sim{}10^6$ to $\sim{}10^8$, for the dimension after reduction is
$\sim{}10^2$ to $\sim{}10^3$.

The $k$-nearest neighbor problem is defined as follows: when a set of
vectors $v_1 \ldots v_n \in \mathbb{R}^d$, distance function $\delta$
and an integer $k$ is given, find $k$ nearest vectors to each $v_i$.  We
propose an effective and scalable algorithm to solve it on multiple
Graphics Processor Units (GPUs). Our algorithm is implemented in
CUDA~\cite{cuda_zone}, which is extension of C language provided by
NVIDIA.

A GPU is a powerful commodity processor.
Although a GPU is originally designed for processing of graphics, the
movement of the GPGPU (General Purpose computing on GPU) has arisen as an
expected breakthrough for a large scale numerical computation.  The
typical characteristic of the GPGPU is highly massive parallelism. A GPU
has hundres of cores, and to
extract its power, it is necessary to run tens of thousand of threads
per unit. Because of that property, a GPU consumes large energy as a unit,
but it is energy effective per FLOPS.

The algorithm of the $k$-nearest neighbor problem is fundamentally a
combination of $N$-body problem and partial sorting. Nyland et
al.~\cite{nyland07} showed an effective algorithm for $N$-body problem
on CUDA. Because dealing with high dimensional vectors, we give some
trick in addition to the known $N$-body algorithm. About sorting,
\cite{cederman08} showed an effective algorithm, but we have employed
another algorithm because we have to sort many arrays at once and we
only need to have top $k$ element not fully sorted data. 

Garcia et al.~\cite{garcia08} showed a GPU algorithm to compute the $k$-nearest
neighbor problem with respect to Kullback-Leibler divergence. Their
algorithm mainly uses a texture memory, which in effect, works as a cache
memory. Its performance largely depends on the cache-hit ratio, and for a
large data, it is likely that a cache miss occurs frequently. On the other
hand, our algorithm utilizes maximal power of coalesced memory access,
so that such loss as a cache miss never happens. Moreover, our algorithm
is effective even for a symmetric distance function and for multiple GPUs.

The rest of this paper is organized as follows. In Sect.~\ref{sec:cuda},
outline of CUDA's programming model is explained. In
Sect.~\ref{sec:description}, we define the problem formally. We give
overview of the algorithm in Sect.~\ref{sec:overall}. In
Sect.~\ref{sec:phase1} and \ref{sec:phase2}, we explain the detail of
each step of the algorithm. In Sect.~\ref{sec:experiment}, we show the
result of experiment. We conclude in Sect.~\ref{sec:conclusion}.

\section{Programming model of CUDA}\label{sec:cuda}
In this section, programming model of CUDA is briefly explained. For
more details of CUDA, refer to \cite{cuda_programming}.
\paragraph{Thread model.}
NVIDIA's recent graphics processor contains hundreds of {\em stream processors}
(SPs). An SP is like a core in a CPU; it can compute
simultaneously. For example, GTX280 has 240 SPs. With such many SPs and
very low cost of context switch, a GPU performs well for tens of thousands
of threads. Threads are divided into thread blocks. Each thread block
can contain at most 1024 threads. A function to synchronize threads in a
block is provided, while there is no such function to synchronize
thread blocks. The only way to synchronize thread blocks is to bring
back the control to the CPU.
\paragraph{Hierarchal memories.}
Before running a GPU, the CPU must explicitly copy data to the GPU's
memory. The memory in GPU to share the data with CPU is called {\em
global memory}. 
A thread block is also a unit to share data. Each thread block has a
memory to share only in the thread block. It is called {\em shared memory}.
The access to the global memory is relatively slow, and usually copying
necessary data to shared memory is better for performance. Although
global memory has some gigabytes, shared memory has only 16KB for each
thread block. Each thread also has a local memory which is called a {\em
register}. The access to a register is fast, but its size is also limited.
\paragraph{Coalesced memory access.}
In CUDA, for example, if successive 16 threads are accessing the
successive 128 bytes in global memory at the same time, the memory
access is coalesced. When a memory access is coalesced, it is done in
only one fetch while otherwise access by 16 threads takes 16 fetches. Hence,
effective utilization of coalesced memory access affects very much the
total performance of an application. The detailed condition about when
memory access can be coalesced is explained in \cite{cuda_programming}.

\section{Description of the problem}\label{sec:description}
The $k$-nearest neighbor problem is described as follows.
\begin{quote}
 Suppose that a set of vectors $v_1, \cdots ,v_n \in \mathbb{R}^d$
 and distance function $\delta : \mathbb{R}^d \times \mathbb{R}^d \to
 \mathbb{R}$ is given. 
 Then output the $k$ nearest vectors to each $v_i$.

 In other words, for each $i$, find a subset of indices
 $\{j_{i1},\ldots,j_{ik}\} \subset \{1,2,\ldots,n\}$ such that 
\[
 \delta(v_i,v_{j_{i1}})\leq \delta(v_i,v_{j_{i2}})\leq \cdots \leq\delta(v_i,v_{j_{ik}})
\]
 and
\[
 \delta(v_i,v_{j_{ik}})\leq \delta(v_i,v_j) \text{ for all } j \not\in 
 \left\{ j_{i1},\ldots,j_{ik} \right\}
\]
\end{quote}

The distance function $\delta$ is arbitrary. Although we use the word
``distance'', it does not necessarily need to satisfy the axiom of
distance. We assume that $\delta$ is {\em cumulatively
computable}. It means $\delta$ can be computed step by step by referring
to each coordinate values. In other words, it is computed with a function $\bar\delta:
\mathbb{R} \times \mathbb{R} \times \mathbb{R} \to \mathbb{R}$ and some
initial value $a_1$ by
$a_i=\bar\delta(u_{i-1},v_{i-1},a_{i-1})$ and $\delta(u,v)=a_{n+1}$.

In this paper, we only discuss the case when $\delta$ is symmetric:
i.e. $\delta(u,v) = \delta(v,u)$. In the symmetric case, we can omit the
half of distance calculations, and consequently, balancing of the
workload becomes more difficult. The algorithm explained in this paper
is easily modified for non-symmetric distance function.

\section{Overview of the algorithm}\label{sec:overall}
Since we have assumed that $\delta$ is symmetric, we only compute
$\delta(v_x,v_y)$ for $x>y$. For an explanation, we depict the whole
problem as a square where the point $(x,y)$ stands for the computation
of $\delta(v_x,v_y)$. The distances to compute is represented by upper
right triangle of the square.

\begin{figure}[t]
\begin{center}
   \includegraphics[height=7cm,keepaspectratio,angle=-90,clip]{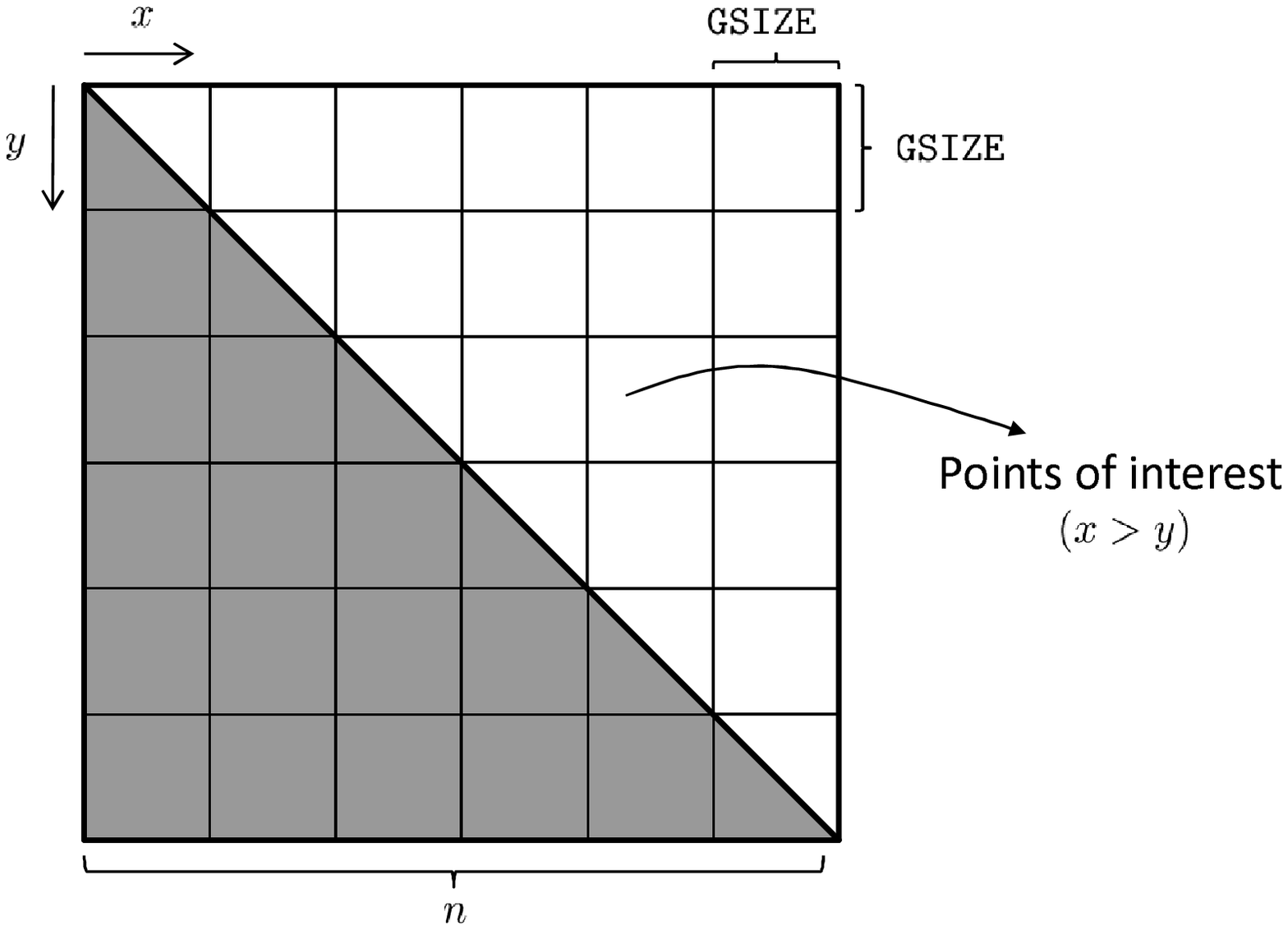}
 \caption{First level division of the problem}\label{fig:superblock}
\end{center}
\end{figure}

\begin{figure}[t]
 \begin{center}
  \includegraphics[height=7cm,keepaspectratio,angle=-90,clip]{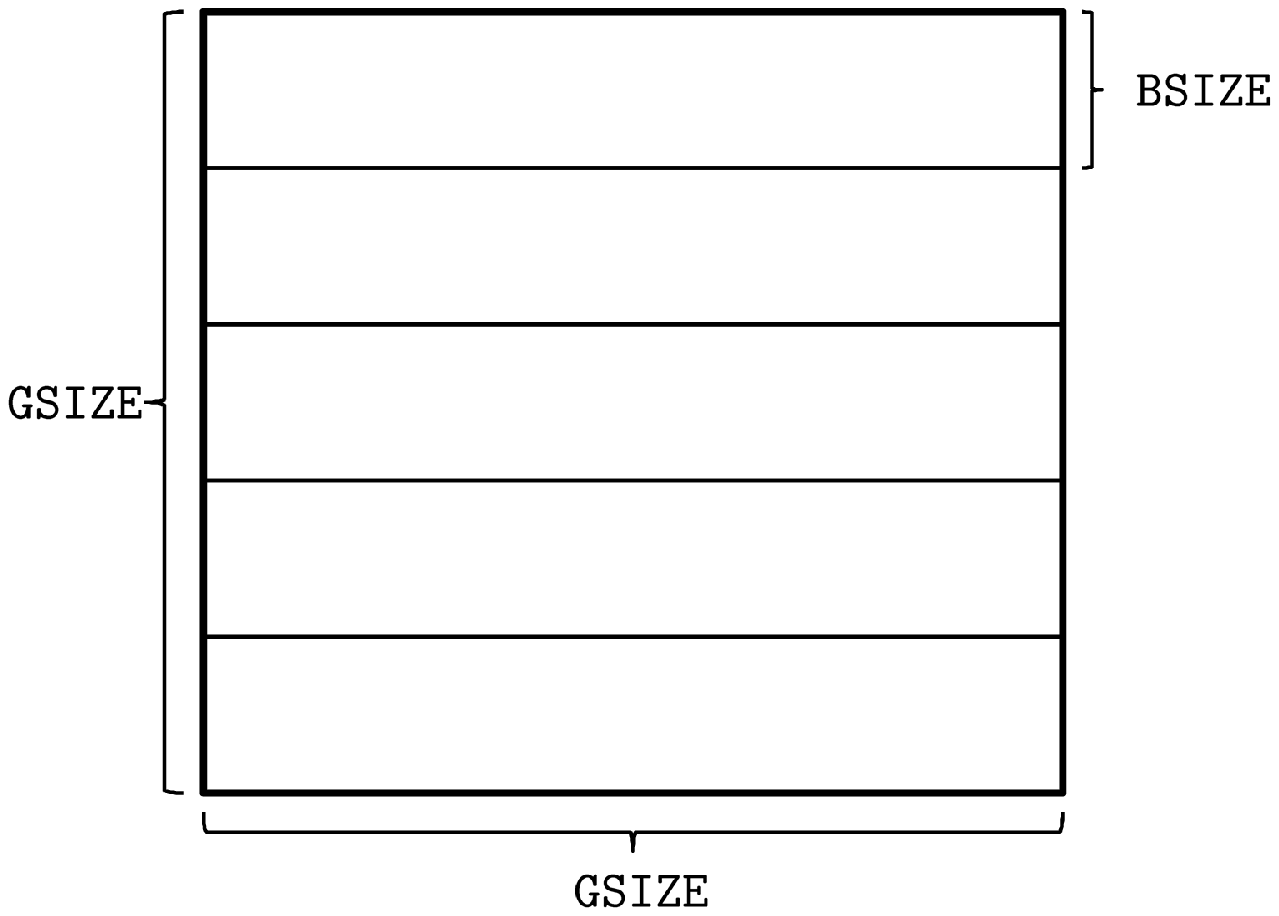}
 \caption{Devision of a grid}\label{fig:blocks}
 \end{center}
\end{figure}

Because of the limitation of the number of threads which can be run at
once on GPU, the problem is divided into a first level blocks. We call
each of them a {\em grid} (Fig.~\ref{fig:superblock}). Each grid is processed in a GPU
at once. A grid can be divided row-wisely into {\em
blocks}, each of which is computed in a thread block. We denote the size
of each side of a grid by $\gsize$. It means the grid $(X,Y)$
stands for the region $\gsize\cdot X \leq x < \gsize \cdot(X+1),\; \gsize\cdot Y
\leq y < \gsize\cdot(Y+1)$. Similarly we denote the size of a block
(i.e.\ the number of rows in a block) by $\bs$ (Fig.~\ref{fig:blocks}).
$\gsize$ is determined depending on $n$ so that the problem can be
devided effectively, while $\bs$ is fixed according to the capability of
CUDA.

\begin{figure}[t]
\begin{center}
  \includegraphics[height=4.5cm,angle=-90,clip]{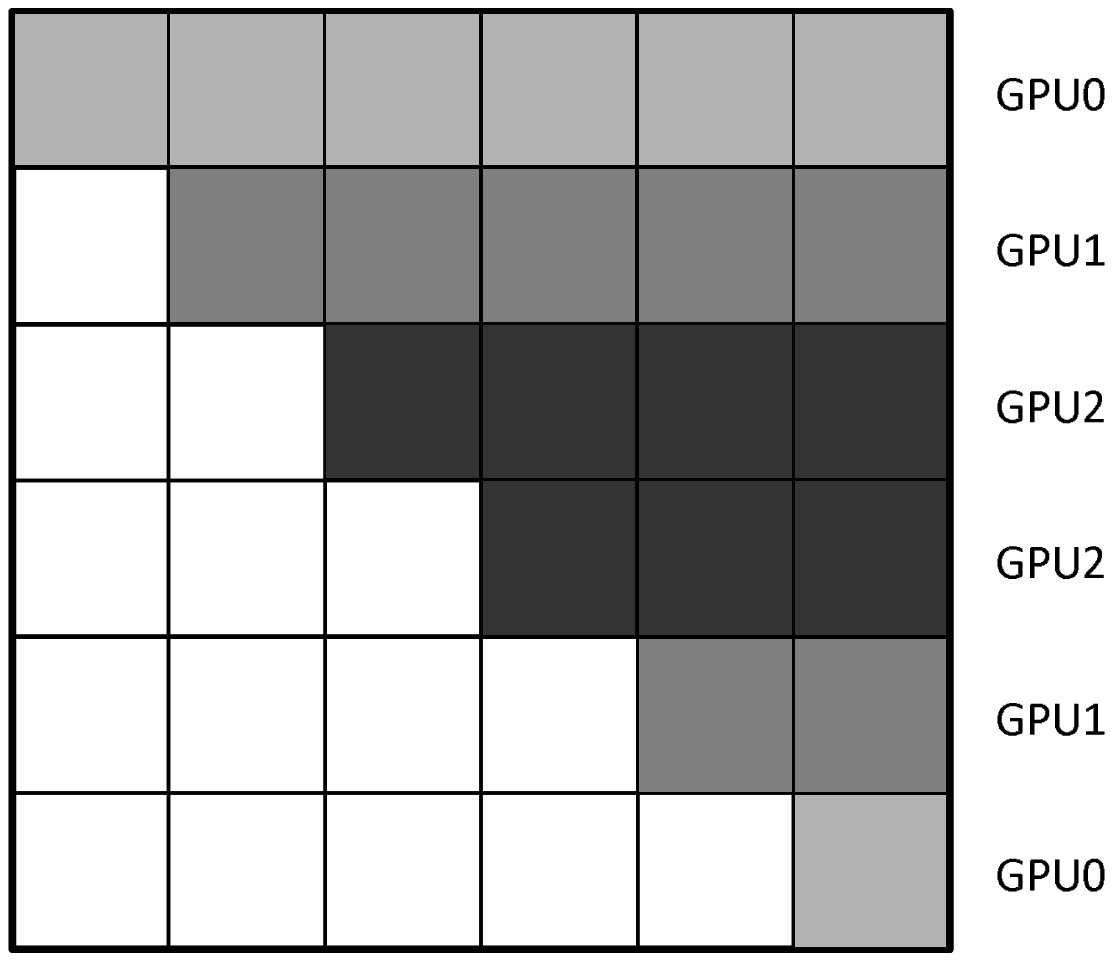}
 \caption{Assignment for GPUs}\label{fig:assignment}
\end{center}
\end{figure}

\begin{figure}[t]
 \centering\includegraphics[scale=0.4,angle=-90,clip]{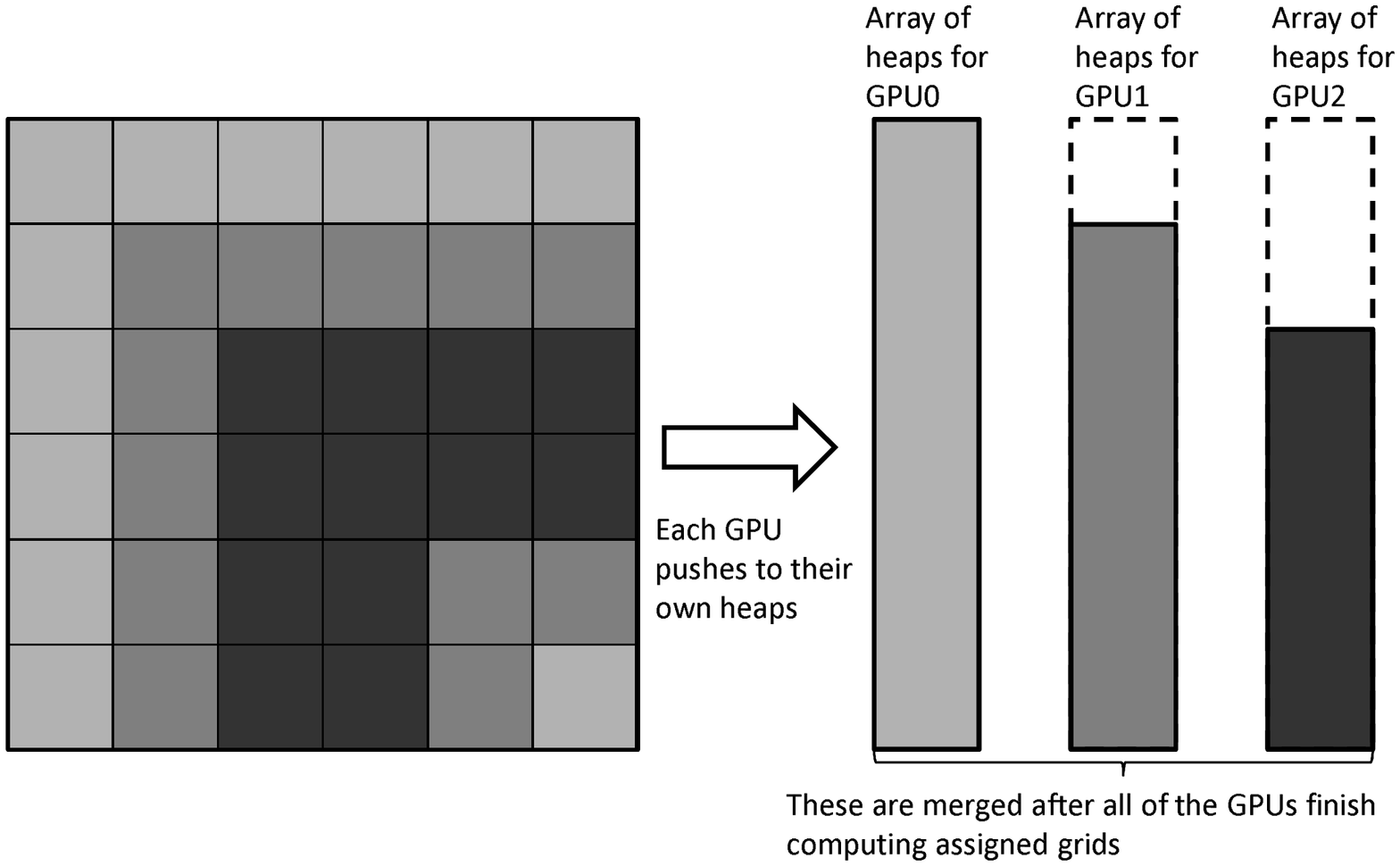}
 \caption{Heaps for GPUs}\label{fig:heaps}
\end{figure}

To balance the workload, we assign GPUs as in
Fig.~\ref{fig:assignment}. In other words, the $i$-th row of grids
is assigned to $j$-th GPU when $i\! \mod 2\cdot\ndev=j$ or $i\! \mod
2\cdot\ndev = 2\cdot\ndev - j -1$, where $\ndev$ means the number of
GPUs available. Here note that although it is enough to compute the
upper-right part of the problem, each GPU virtually compute the mirror
side of the assigned part (see also Fig.~\ref{fig:heaps})

To keep the $k$-nearest vectors, we use a heap structure. The heap has
at most $k$ elements and is in descending order, so that the $k$-th
smallest element can be found in $O(1)$. Moreover, each GPU keeps their
own heaps to avoid a costly synchronization (Fig.~\ref{fig:heaps}). It
means each GPU has $n$ heaps which stores the $k$ nearest elements
computed by itself. At the last phase, the heaps of
different GPUs are merged in CPU.

Thus the outline of the algorithm is shown in Fig.~\ref{alg:overall}.
In this algorithm, the calclation of distances is explained in
Sect.~\ref{sec:phase1}, and how to push the distances to the heaps is
described in Sect.~\ref{sec:phase2}.
\begin{figure}[h]
\begin{tabbing}
 \phantom{X}\=\phantom{X}\=\phantom{X}\=\phantom{X}\=\phantom{X}\=\phantom{X}\=\kill
 {\bf procedure} ThreadMain($n$,$d$,$\left\{v_i\right\}$)\\
 \> $\mathtt{nGrids}\gets \lfloor (n-1)/ \gsize\rfloor +1$\\
 \> Prepare the heaps $\left\{ h_i \right\}_{i=0}^{n-1}$\\
 \> {\bf for} $Y:=0$ {\bf to} $\mathtt{nGrids}-1$ {\bf do}\\
 \>\> {\bf for} $X:=0$ {\bf to} $\mathtt{nGrids}$-1 {\bf do}\\
 \>\>\> {\bf if} $i\! \mod 2\cdot\ndev=j$ \\
 \>\>\>\>\>or $i\! \mod 2\cdot\ndev$ \\
 \>\>\>\>\> $= 2\cdot\ndev - j -1$ {\bf then}\\
 \>\>\>\> Calculate the distances for the grid
 $(X,Y)$\\
 \>\>\>\> Push the $i$-th row of distances\\
 \>\>\>\>\>\>to $h_i$ for the grids $(X,Y)$ and $(Y,X)$\\
 \>\>\> {\bf end if}\\
 \>\> {\bf end for}\\
 \> {\bf end for}\\
 {\bf end procedure}
\end{tabbing}
 \caption{Overall algorithm: each GPU is assigned to CPU
 thread and its thread id is given by $tid$}\label{alg:overall}
\end{figure}
\section{Phase 1: calculation of distances}\label{sec:phase1}
Basically the framework of the process to compute the distances of
vectors is the same as the algorithm of $N$-body problem written in
\cite{nyland07}. A grid is row-wisely devided into blocks, and each
block is assigned to a thread block. Each thread corresponds to a
row. A block first copies a fixed number (which we denote by $\cx$) of
columns to the shared memory. Then compute the distances.

\begin{figure}[h]
\begin{center}
  \includegraphics[scale=0.4,angle=-90,clip]{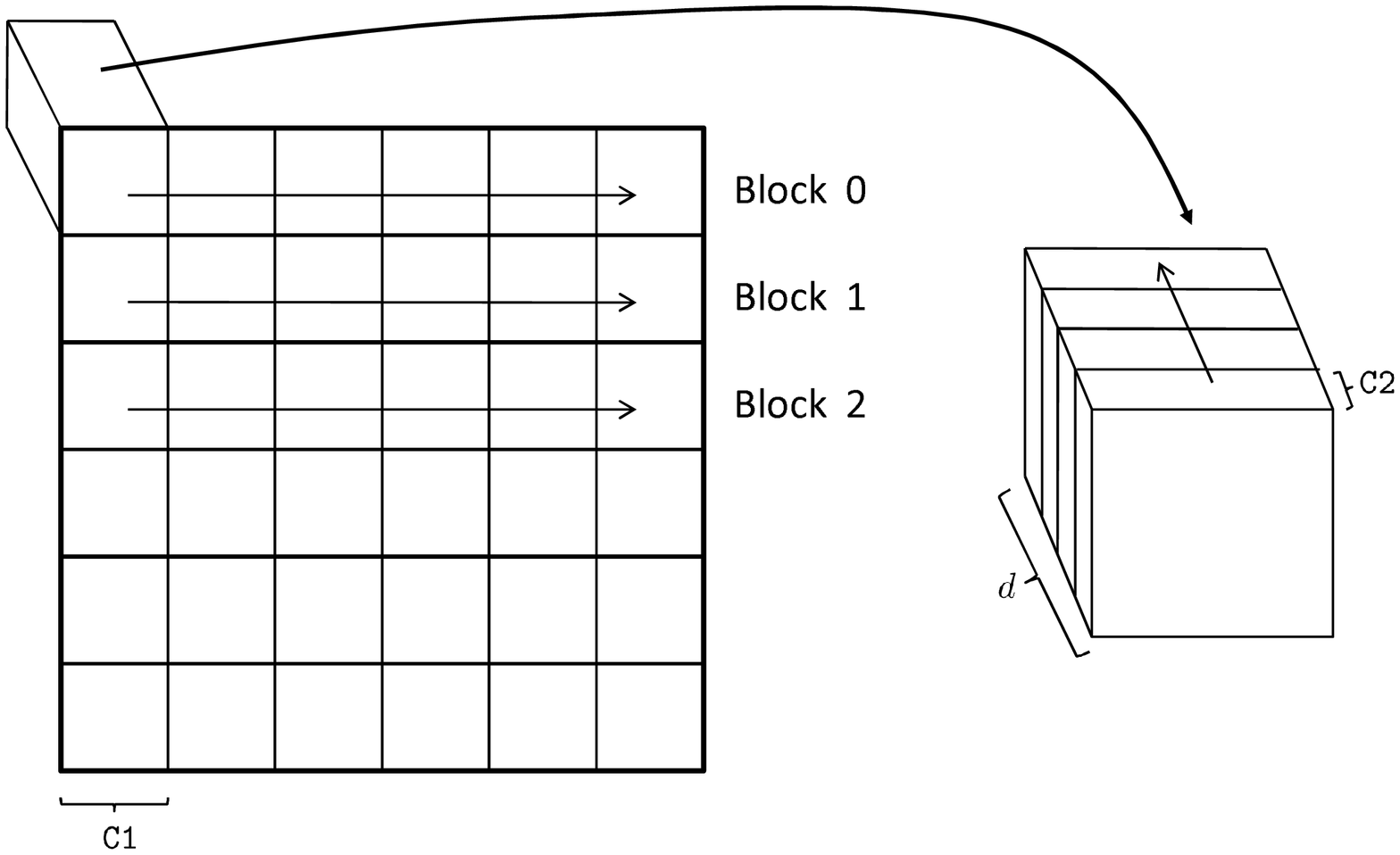}
\end{center}
\caption{Illustration of the algorithm to compute the distaces of $d$
 dimensional vectors}\label{fig:nbody}
\end{figure}

However, in our problem, since the dimension $d$ is large, it is not
possible to copy all the coordinate data to the shared memory even for a
small $\cx$. Hence, a thread iteratively reads a fixed number $\cy$ of
coordinate values of corresponding vectors. In other words, if $v_i$ is
expressed as $(v_i^{(0)}, \ldots, v_i^{(d-1)})$, then $v_i^{(j\cdot
\cy)},\ldots,v_i^{(j\cdot (\cy+1)-1)}$ are read in $j$-th iteration
(Fig.~\ref{fig:nbody}). If a vector is expressed by single precision
numbers, $\cy$ must be a multiple of 32 to utilize full power of
coalesced memory accesses.

The algorithm to calculate the distaces for a given grid is shown
in Fig.~\ref{alg:calcdist}. Here, the arguments $n_1$, $\left\{v_{1i}\right\}_{n=0}^{n_1-1}$,
$n_2$, and $\left\{v_{2i}\right\}_{i=0}^{n_2-1}$ are given as re-indexed $\left\{v_i\right\}$
so that this procedure can calculate for the assigned grid. The index
for the block is expressed by $bid$, and each block has $\bs \times \cx$
threads. Each thread is indexed by $(tx,ty)$
\begin{figure}[h]
\begin{tabbing}
 \phantom{X}\=\phantom{X}\=\phantom{X}\=\phantom{X}\=\phantom{X}\=\phantom{X}\=\kill
 {\bf procedure} CalcDistances($d$,$n_1$,$\left\{v_{1i}\right\}_{i=0}^{n_1-1}$,
  $n_2$,  $\left\{v_{2i}\right\}_{i=0}^{n_2-1}$)\\
 \> $bx\gets 0$\\
 \> Prepare the shared memory to store the distances\\
 \> {\bf while} $bx\cdot \cx <n_1$ {\bf do}\\
 \>\> $l\gets 0$\\
 \>\> {\bf while} $l <d$\\
 \>\>\> Copy $v_{1i}^{(k)}, v_{2j}^{(k)}$\\
 \>\>\>\>\>($bx\leq i<bx+\cy$,\\
 \>\>\>\>\>$bid\cdot\bs\leq j<(bid+1)\cdot \bs$,\\
 \>\>\>\>\>$l\leq k < l+\cy$) to the shared memory\\
 \>\>\> Calculate cumulatively all the combinations of\\ 
 \>\>\>\>\> $v_{1i}$ and $v_{2i}$ which are in the shared memory \\
 \>\>\>\>\> and store it in a local resister $dist$\\
 \>\>\> $l\gets l+\cy$\\
 \>\>{\bf end while}\\
 \>\> Store the resulting distance $dist$ in the global memory\\
 \>\> $bx \gets bx+\cx$\\
 \>{\bf end while}\\
 {\bf end procedure}
\end{tabbing}
 \caption{Algorithm for calculation of distances: for simplicity, it
 is assumed that $n_1$ is multiple of $\cx$ and $d$ is multiple of
 $\cy$}\label{alg:calcdist}
\end{figure}
\section{Phase 2: taking $k$ smallest elements}\label{sec:phase2}
In the second phase, each thread block is assigned to each row. The
smallest $k$ distances are computed by parallel processing of threads in
the block. If the number of thread in a block is denoted by
$\mathtt{nThreads}$, each thread read distances skipping
$\mathtt{nThreads}$, so that memory access is coalesced. A thread check
if the element is smaller than current $k$-th largest element in the
heap, and store it in the local buffer if so. This is because $k$ is
relatively small than $n$ and it is likely that only a few elements is
stored in the local buffer. Because of this mechanism, the waiting time
is shortened even though when pushing to the heap, the threads must
be synchronized.

The algorithm is shown in Fig.~\ref{alg:ksmallest}. Here, the index for the
block and thread is denoted by $bid$ and $tid$ respectively, and
$\mathtt{buffer}$ is thread-local array and its size is
$\mathtt{bufsize}$.
\begin{figure}[h]
\begin{tabbing}
  \phantom{X}\=\phantom{X}\=\phantom{X}\=\phantom{X}\=\phantom{X}\=\phantom{X}\=\kill
 {\bf procedure} KSmallest ($k$,$n$,$m$,$\{h_i\}$: heaps, \\
 \hspace{5cm}$\{a_{ij}\}_{0\leq  i<m-1, 0\leq j<n-1}$)\\
  \>{\bf for} $i:=tid$ {\bf to} $n-1$ {\bf step}
 $\mathtt{nThreads}\cdot\mathtt{bufsize}$ {\bf do}\\
  \>\>{\bf for} $j:=0$ {\bf to} $\mathtt{nThreads}\cdot\mathtt{bufsize}$
  {\bf step} $\mathtt{bufsize}$ {\bf do}\\
  \>\>\>{\bf for} $l:=0$ {\bf to} $\mathtt{bufsize}$ {\bf do}\\
  \>\>\>\> $\nu\gets i\cdot\mathtt{nthreads}\cdot\mathtt{bufsize}  + j
  + l$\\
  \>\>\>\>{\bf if} $a_{bid,\nu}$ is smaller than top of the heap $h_{bid}$ {\bf then}\\
  \>\>\>\>\>Store $a_{bid,\nu}$ to $\mathtt{buffer}$\\
  \>\>\>\>{\bf end if}\\
  \>\>\> {\bf end for}\\
  \>\>\> Push elements of $\mathtt{buffer}$ \\
 \>\>\>\>\>to $h_{bid}$ (blocking other
  threads)\\
  \>\>{\bf end for}\\
  \>{\bf end for}\\
 {\bf end procedure}
\end{tabbing}
 \caption{Algorithm to get $k$-smallest numbers from multiple
 arrays}\label{alg:ksmallest}
\end{figure}
\section{Experiment}\label{sec:experiment}

We experimented our algorithm on two GTX280's and one GTX280. For a
comparison, we also implemented CPU version and experimented it on Intel
i7 920 (2.67GHz). GTX280 is one of the latest NVIDIA's graphics chips. The
algorithm experimented on the CPU is a simple one: it calculates each
$\delta(v_x,v_y)\ (x>y)$ and pushes it to the corresponding heaps. Note that
although Intel i7 has four cores with hyperthreading capability, we only
worked on serial algorithm, i.e. it only uses one core.

The distance employed here is Hellinger distance, which often used in
the context of statistics. Hellinger distance for two vectors $u$ and
$v$ is defined as:
\begin{equation}
 \sum_i \left(
\sqrt{u^{(i)}} - \sqrt{v^{(i)}}
\right)
\end{equation}

The result of the experiment for various $n$ is shown in Table
\ref{tab:perf}. The other parameters are set as $k=100$ and $d=256$; and
the data is generated randomly.  It shows that for a large problem, our
algorithm work well from the viewpoint of parallelism of GPUs. Moreover,
it also tells the GPUs substantially outperforms the CPU; for a large
problem, two GPU implementation is more than 330 times faster than the
CPU.
\begin{table}[t]
\caption{Elapse time for $k$-nearest neighbor problem (sec)}\label{tab:perf}
\begin{center}
  \begin{tabular}{c|rrrr}
 $n$ &10000 &20000 &40000 &80000\\
 \hline
 $2\times$ GTX280 (a)& 1.8& 5.7& 17.7& 68.6\\
 $1\times$ GTX280 (b)& 2.7& 8.6& 34.1& 131.8\\
 i7 920 (CPU) (c)& 354.2& 1419.0& 5680.7& 22756.9\\
 \hline
   (c)/(a) & 196.7& 248.9& 320.9& 331.7\\
   (c)/(b) & 131.1& 173.3& 166.5& 172.6\\
   (b)/(a) & 1.50& 1.51& 1.92& 1.92
 \end{tabular}
\end{center}
\end{table}

\section{Conclusion}
\label{sec:conclusion}
We introduced an effective algorithm for $k$-nearest neighbor problem
which works on multiple GPUs. By
an experiment, we have shown that it runs more than 330 times faster than
an implementation on a single core of an up-to-date CPU. We have also shown
that the algorithm is effective from the viewpoint of parallelism of
GPUs. That is because 1) there is no synchronization between GPUs until
the very end of the process and 2) the workload is well balanced.

Our algorithm includes simultaneous partial sort of multiple arrays. It
minimizes the inter-thread synchronization utilizing the fact that if
$k\ll n$, most of the data are discarded. About this part of algorithm, we
have achieved a good performance but still there is a room for
improvement because it uses arrays in a local scope which are stored in a
slow global memory in effect. To improve the performance of the
simultaneous partial sort is our ongoing work, and we believe this problem
alone is also important because it can be applied to other problems.

\section*{Acknowledgment}
The authors would like to thank Khan Vo Duc of NVIDIA for giving us
a helpful advice about an early version of this paper.

\bibliographystyle{IEEEtran}
\bibliography{cuda}

\begin{thebibliography}{10}

\bibitem{cuda_zone}
NVIDIA:
\newblock {CUDA} {Zone}.
\newblock \url{http://www.nvidia.com/object/cuda_home.html}

\bibitem{amazon}
Amazon.com.
\newblock \url{http://www.amazon.com}

\bibitem{netflix}
Netflix.
\newblock \url{http://www.netflix.com}

\bibitem{brand03}
Brand, M.:
\newblock Fast online {SVD} revisions for lightweight recommender systems.
\newblock In: In SIAM International Conference on Data Mining. (2003)

\bibitem{blei03}
Blei, D.M., Ng, A.Y., Jordan, M.I., Lafferty, J.:
\newblock Latent dirichlet allocation.
\newblock Journal of Machine Learning Research \textbf{3} (2003)  2003

\bibitem{das07}
Das, A.S., Datar, M., Garg, A., Rajaram, S.:
\newblock Google news personalization: scalable online collaborative filtering.
\newblock In: WWW '07: Proceedings of the 16th international conference on
  World Wide Web, New York, NY, USA, ACM (2007)  271--280

\bibitem{indyk98}
Indyk, P., Miller, W.:
\newblock Approximate nearest neighbors: towards removing the curse of
  dimensionality.
\newblock In: In proceedings of the 1998 Symposyum on Theory of Computing.
  (1998)

\bibitem{nyland07}
Nyland, L., Harris, M., Prins, J.:
\newblock Fast $n$-body simulation with {CUDA}.
\newblock In: GPU Gems III.
\newblock NVIDIA (2007)  677--695

\bibitem{cederman08}
Cederman, D., Tsigas, P.:
\newblock A practical quicksort algorithm for graphics processors.
\newblock In: ESA '08: Proceedings of the 16th annual European symposium on
  Algorithms, Berlin, Heidelberg, Springer-Verlag (2008)  246--258

\bibitem{cuda_programming}
NVIDIA:
\newblock {CUDA} 2.1 programming guide.
\newblock \url{http://www.nvidia.com/object/cuda_develop.html} (2008)

\end{thebibliography}
\end{document}